\documentclass{Interspeech2024}
\usepackage{rotating}
\usepackage{booktabs}
\usepackage{xfrac}
\usepackage{xcolor}
\usepackage{tablefootnote}
\usepackage{multirow}
\usepackage{cleveref}
\usepackage{adjustbox}
\usepackage{enumitem}
\usepackage{graphicx}
\usepackage{float}
\usepackage{xcolor}
\usepackage{afterpage}
\usepackage{adjustbox}
\definecolor{xiaomi_gray}{HTML}{A9A9A9}

\interspeechcameraready

\title{Bridging Language Gaps in Audio-Text Retrieval}

\name{Zhiyong}{Yan}
\name{Heinrich}{Dinkel}
\name{Yongqing}{Wang}
\name{Jizhong}{Liu}
\name{Junbo}{Zhang}
\name{Yujun}{Wang}
\name{Bin}{Wang}

\address{
  AI Lab, Xiaomi Corporation, China}
\email{\{yanzhiyong, dinkelheinrich, wangyongqing3, liujizhong1, zhangjunbo1, wangyujun, wangbin11\}@xiaomi.com}


\begin{document}

\maketitle

\begin{abstract}
    
Audio-text retrieval is a challenging task, requiring the search for an audio clip or a text caption within a database. 
The predominant focus of existing research on English descriptions poses a limitation on the applicability of such models, given the abundance of non-English content in real-world data.
To address these linguistic disparities, we propose a language enhancement (LE), using a multilingual text encoder (SONAR) to encode the text data with language-specific information.
Additionally, we optimize the audio encoder through the application of consistent ensemble distillation (CED), enhancing support for variable-length audio-text retrieval. 
Our methodology excels in English audio-text retrieval, demonstrating state-of-the-art (SOTA) performance on commonly used datasets such as AudioCaps and Clotho. 
Simultaneously, the approach exhibits proficiency in retrieving content in seven other languages with only 10\% of additional language-enhanced training data, yielding promising results.

\end{abstract}

\section{Introduction}
Audio-text retrieval, requiring the search for an audio clip or a caption within a database, based on a query from another modality, has seen significant advancements and applications in recent years. 
The integration of audio and text has facilitated various applications such as content-based audio search~\cite{556537}, and multimedia information retrieval.
Audio-text retrieval is also one of the tasks featured in the Detection and Classification of Acoustic Scenes and Events (DCASE) competition~\cite{7100934}. 
A widely adopted technique in this field is Contrastive Language-Audio Pretraining (CLAP)~\cite{elizalde2022clap,guzhov2021audioclip,wu2022wav2clip} inspired by CLIP~\cite{radford2021learning, yu2022coca, li2022blip}, which has demonstrated remarkable success in learning robust representations for audio-text retrieval tasks.

One significant limitation of current audio-text retrieval systems is their focus on monolingual retrieval, often restricted to single-language queries such as English. 
While there are datasets with non-English captions, such as~\cite{wu2019audio}, these datasets are small and often contain other errors such as imprecise annotations.
However, advancements in multilingual text translation technology and the growing availability of open-source tools, such as OpusMT~\cite{Tiedemann2020OPUSMTB} and NLLB~\cite{nllbteam2022language} have made it feasible to perform large-scale multilingual audio-text retrieval. 
This is achieved by leveraging automatic translation for data augmentation. 
Research in the multilingual AAC~\cite{cousin2023multilingual} has validated the viability of this method. 
Their proposed solution, however, suffers from limited language scalability, noting a lack of comprehensive evaluation regarding their performance across various languages. 


%

In the realm of existing audio-text retrieval systems, various audio encoders have been employed, each with its strengths and limitations. 
HTSAT~\cite{chen2022htsat,mei2023wavcaps}, Audio-MAE in FLAP~\cite{yeh2023flap} and Cacophony~\cite{zhu2024cacophony} offer a promising alternative, particularly in capturing long-range dependencies in audio sequences. 
However, all these encoders struggle modeling variable-length audio segments.
These limitations highlight the need for novel approaches to enhance the performance and adaptability of audio encoders in multilingual audio-text retrieval systems.

To address these challenges, this paper presents two primary contributions.
\begin{itemize}
  \item We incorporate language enhancement (LE) into retrieval tasks, employing a multilingual text encoder. 
  SONAR~\cite{Duquenne:2023:sonar_arxiv}, featuring a comprehensive suite of speech and text encoders and decoders, is one of the eligible candidates. 
  We utilize its text-decoder for the generation of multilingual training data and its text encoder for multilingual text encoding, thereby bridging language gaps in the field.
  \item We optimize the audio encoder through the application of CED~\cite{dinkel2023ced} to overcome performance limitations when dealing with variable-length audio-text retrieval.
\end{itemize}
The experimental results indicate that a moderate portion of multilingual training serves as a form of data augmentation for standalone English audio-text retrieval, leading to a significant improvement in performance. 
We also achieve state-of-the-art (SOTA) results on widely used datasets such as AudioCaps and Clotho in English audio-text retrieval, demonstrating proficiency in retrieving content across seven additional languages.

\begin{figure*}[t]
  \centering
  \begin{adjustbox}{max width=\textwidth}
    \includegraphics[height=0.5\textheight,keepaspectratio]{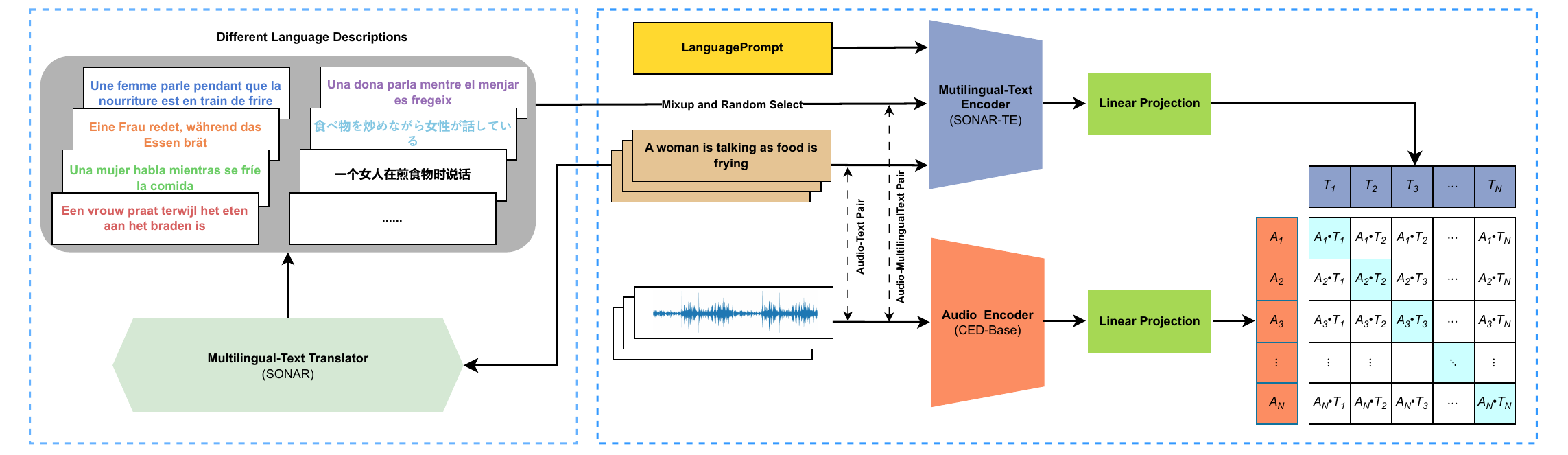}
  \end{adjustbox}
  \caption{The proposed multilingual audio-text retrieval framework. 
  We first generate multilingual text descriptions of the training data using the SONAR text decoder, displayed on the left.
  Then we train a multilingual audio-retrival model based on CLAP, which can be seen on the right.
  Models are evaluated by translating test-captions using ChatGPT.
  }
  \label{fig:ml-clap}
\end{figure*}

\section{Methodology}
The details of the multilingual audio-text retrieval are illustrated in \Cref{fig:ml-clap}. 
It consists of two primary components: the offline preparation of multilingual data and the model training framework.

\paragraph*{Multilingual Data Preparation}

A multilingual text translator is employed to translate the English descriptions from the training set into seven additional languages. 
Considering that each audio clip in the Clotho training dataset \cite{drossos2020clotho} is associated with multiple captions, a single one is randomly selected for each language translation. 
Additionally, each translated caption is annotated with a language prompt, such as {eng, fra, deu} and so forth.

During each training epoch, a subset of the multilingual descriptions is sampled from the translated text data randomly and added to the training set.
These samples are then combined with the original English descriptions to form multilingual audio-text pairs using the same audio. The performance of the subset sampled at different percentages is shown in \Cref{ssec:ablation}.


\paragraph*{Training Framework}
The essence of the audio-text retrieval task lies in comparing the similarity between the audio and text modalities, with CLAP~\cite{elizalde2022clap} being one of the most commonly used techniques to achieve this.
It employs a bi-encoder architecture comprising an audio encoder ${E_A}$, a text encoder ${E_T}$, and a cross-modal matching module~\cite{sun2023leveraging}. 
These encoders transform an audio-text pair $(\mathcal{A}, \mathcal{T})$ into an embedding pair $(e_a, e_t)$, which are subsequently linked in a joint cross-modal space using linear projections. This space is trained through contrastive learning, leveraging the (dis)similarity of audio and text pairs within a batch~\cite{mei2023wavcaps}. 

Similar methodologies are employed in the multilingual audio-text retrieval task.
The audio-text pairs spanning multiple languages are fed into a shared text encoder that facilitates multilingual text encoding, bolstered by the addition of a language prompt.
This process is termed as language enhancement (LE). We also introduce the concept of mixture LE, where the audio-text pairs encompass all seven additional languages detailed in this paper.
We slightly modify the text encoder ${\text{E}_T}$ to ${\text{E}_{MT}}$, denoting its adaptation for multilingual text processing:
\begin{equation}
\begin{aligned}
e_a&=\text{E}_A(\mathcal{A} ), \\
e_t&=\text{E}_{MT}(\mathcal{T} ),\\
a&=\text{Project}_A(e_a), \\
t&=\text{Project}_{MT}(e_t).
\end{aligned}
\end{equation}
The similarity score (cosine similarity in this system) between $a$ and $t$ is computed as:
\begin{equation}
s_{A \sim MT}=\frac{a_p\cdot t_p^T} {||a_p|| \cdot ||t_p||}
\end{equation}

The InfoNCE loss~\cite{oord2019representation} is adopted as the loss function. 
This contrastive training loss between the similarity scores and the ground truth labels is calculated as follows:
\begin{equation}
\begin{aligned}
\mathcal{L} _{i}^{A\longrightarrow MT}&=- \log_{}{\frac{\exp (s_{{A\sim MT}}(i,i)/\tau )}{ {\textstyle \sum_{j=1}^{N}\exp (s_{{A\sim MT}}(i,j)/\tau )} } }, \\
\mathcal{L} _{i}^{MT\longrightarrow A}&=- \log_{}{\frac{\exp (s_{{A\sim MT}}(i,i)/\tau )}{ {\textstyle \sum_{j=1}^{N}\exp (s_{{A\sim MT}}(j,i)/\tau )} } }, \\
\mathcal{L}&=\frac{1}{N} \sum_{i=1}^{N} (\mathcal{L} _{i}^{A\longrightarrow MT}+\mathcal{L} _{i}^{MT\longrightarrow A}),
\end{aligned}
\end{equation}
where $\tau$ is a temperature hyper-parameter.

In our work, the model architecture primarily consists of SONAR-TE (SONAR text encoder) as the text encoder ${\text{E}_{MT}}$ and CED as the audio encoder ${\text{E}_A}$.

\section{Experiments}
\subsection{Dataset}
In our experiments, we use the AudioCaps~\cite{audiocaps} and Clotho~\cite{drossos2020clotho} and WavCaps~\cite{mei2023wavcaps} datasets. 
The AudioCaps contains about 49,000 audio samples, each lasting around 10 seconds. 
Each audio is associated with a single sentence in the training set, while in the validation and test sets, each audio has five annotated sentences.
The Clotho consists of 6,974 audio samples, ranging from 15 to 30 seconds in length, and each audio sample is annotated with five sentences. 
The dataset is divided into 3,839 training samples, 1,045 validation samples, and 1,045 test samples. 
WavCaps is a large-scale weakly-labelled audio captioning dataset, comprising approximately 400k audio clips with paired captions.
Its main data sources include four parts: FreeSound, BBC Sound Effects, SoundBible, and the Strongly-Labelled Subset of AudioSet.

Furthermore, we perform automatic translation of the training datasets from the AudioCaps and Clotho datasets into seven languages for the training of multilingual audio-text retrieval, utilizing a multilingual text translator based on the SONAR text decoder.


\subsection{Models}

\paragraph*{Audio Encoder}
For the Audio Encoder, we use the recently introduced CED-Base model~\cite{dinkel2023ced}.
CED-Base is a standard 86 M parameter vision transformer that has been trained on Audioset \cite{gemmeke2017audio} via knowledge distillation from a large teacher ensemble. 
The model uses 64-dimensional Mel-spectrograms as inputs computed from a 16 kHz signal.
Then it extracts non-overlapping $16 \times 16$ patches from the Mel-spectrogram, which results in $4 \times 62 = 248$ patches over an input of 10s. In our experiments, applying a patch dropout of 25\% on both frequency and time patches yields better results while also accelerating the training speed.

\paragraph*{Text Encoder}
The core of multilingual audio-text retrieval lies in the text encoder's capacity to process multilingual texts. 
In this study, we exclusively use SONAR-TE~\cite{Duquenne:2023:sonar_arxiv}.
SONAR-TE extracts a single vector bottleneck to represent the entire text, without utilizing token-level cross-attention found in standard sequence-to-sequence MT architectures.
The fixed-size text representation is computed by pooling the token-level outputs of the encoder. 
In subsequent sections, SONAR simply represents the text encoder.

\subsection{Setup}
\label{ssec:setup}


Our training dataset is divided into two types: small and large, where the small contains AudioCaps and Clotho and the large contains WavCaps, AudioCaps, and Clotho.
We use ChatGPT 3.5 to translate the captions of the AudioCaps and Clotho test sets into seven different languages, including French (fre), German (deu), Spanish (spa), Dutch (nld), Catalan (cat), Japanese (jpn), and Chinese (zho). 
These serve as the test sets for the multilingual audio-text retrieval task.

This paper's experiments are organized as follows.
We first compare the impact of audio encoders by training on the small dataset.
Next, we train various models using different LE on the small dataset, evaluating their impact on the English test sets, and simultaneously implement multilingual audio-text retrieval with the mixture LE.
After pretraining on the large dataset, the models are fine-tuned on the AudioCaps and Clotho datasets, incorporating the proposed mixture LE approach. 

All models are trained for 20 epochs with a batch size of 128 and a learning rate of $5\times10^{-5}$ using the Adam optimizer, except during fine-tuning where a smaller learning rate $5\times10^{-6}$ is needed.
The temperature hyperparameter $\tau$ is set to 0.07 for all settings. The source code is publicly available\footnote{\url{https://github.com/zyyan4/ml-clap}}.

\subsection{Evaluation metrics}

In audio-text retrieval tasks, the evaluation of model performance relies on the recall at rank k (R@k). 
For a query, R@k is 1 if the target value item appears in the top k retrieved items, otherwise
0. 
The final R@k is averaged across the dataset~\cite{mei2023wavcaps}.
Furthermore, this study introduces the mean average precision at rank 10 (mAP10) metric to offer a more comprehensive comparison of the model's performance variations.

\begin{table}[bt]
\centering
\caption{Comparison of audio encoder based on SONAR-SE and CED respectively.}
\resizebox{\columnwidth}{!}{
\begin{tabular}{ccc|cc|cc|cc}
\toprule
\multirow{4}{*}{Audio Encoder} & \multicolumn{4}{c}{AudioCaps} & \multicolumn{4}{c}{Clotho} \\
\cmidrule(lr){2-9}
& \multicolumn{2}{c}{Audio-to-Text} & \multicolumn{2}{c}{Text-to-Audio} & \multicolumn{2}{c}{Audio-to-Text} & \multicolumn{2}{c}{Text-to-Audio} \\
\cmidrule(lr){2-9}
& R@1 & mAP10 & R@1 & mAP10 & R@1 & mAP10 & R@1 & mAP10 \\
\midrule
SONAR-SE & 0.0 & 0.0 & 14.3 & 31.7 & 0.0 & 0.0 & 12.5 & 32.9 \\
CED & 50.1 & 37.4 & 40.7 & 55.8 & 21.1 & 15.5 & 18.8 & 29.8 \\
\bottomrule
\end{tabular}
}
\label{tab:audio_encoder}
\end{table}

\section{Results}

\subsection{Audio encoder comparison}
Since SONAR itself features a speech encoder (SONAR-SE), this experiment assesses whether this encoder is suited as an audio encoder for retrieval.
The results in \Cref{tab:audio_encoder} indicate that SONAR-SE is not suitable as an audio encoder for audio-text retrieval tasks. 
SONAR-SE shows a strong correlation between speech and text, whereas general audio used in this work exhibits a different pattern.
Therefore we use CED as our default audio encoder in the rest of the paper.

\subsection{Evaluation of LE on English}
\label{ssec:evaluate_LE}
In this section, we demonstrate the impact of enhancing English retrieval through the different LE, as shown in Table \ref{tab:enhancing_text}. 
LE notably enhances the performance of English retrieval, with improvements of up to about 3\% across multiple different languages absolute for both R@1 and mAP10 metrics. 
Notably, R@1 in Audio-to-Text on AudioCaps achieves over a 6\% absolute improvement by using LE with Catalan.
Further, when training with mixture LE, a remarkable performance improvement is also seen.
%
%

\begin{table}[tb]
\centering
\caption{Performance impact of LE on the (original) {English test sets}, where ``baseline'' indicates no enhancement and ``mixture'' denotes the proposed approach.}
\resizebox{\columnwidth}{!}{
\begin{tabular}{ccc|cc|cc|cc}
\toprule
\multirow{4}{*}{LE} & \multicolumn{4}{c}{AudioCaps} & \multicolumn{4}{c}{Clotho} \\
\cmidrule(lr){2-9}
& \multicolumn{2}{c}{Audio-to-Text} & \multicolumn{2}{c}{Text-to-Audio} & \multicolumn{2}{c}{Audio-to-Text} & \multicolumn{2}{c}{Text-to-Audio} \\
\cmidrule(lr){2-9}
 & R@1 & mAP10 & R@1 & mAP10 & R@1 & mAP10 & R@1 & mAP10 \\
\midrule
baseline & 50.1 & 37.4 & 40.7 & 55.8 & 21.1 & 15.5 & 18.8 & 29.8 \\
\hline
 fra & 53.8 & 38.9 & 42.1 & 57.2 & 24.3 & 16.4 & 19.8 & 30.7 \\
 deu & 53.1 & 39.6 & 42.3 & 57.6 & 24.3 & 16.7 & 20.1 & 30.8 \\
 spa & 52.3 & 39.6 & 43.2 & 57.8 & 25.5 & 16.7 & 19.8 & 30.9 \\
 nld & 52.7 & 39.5 & 42.5 & 57.6 & 25.2 & 16.7 & 19.3 & 30.5 \\
 cat & 56.3 & 40.3 & 43.7 & 58.4 & 24.0 & 16.6 & 19.8 & 30.8 \\
 jpn & 54.1 & 39.8 & 43.4 & 58.2 & 24.5 & 16.7 & 19.7 & 30.8 \\
 zho & 52.5 & 39.6 & 42.4 & 57.4 & 23.3 & 16.4 & 19.2 & 30.6 \\
 \hline
 mixture & 53.8 & 39.6 & 42.4 & 57.4 & 24.6 & 16.4 & 18.9 & 29.9 \\
\bottomrule
\end{tabular}
}
\label{tab:enhancing_text}
\end{table}

\begin{figure}[tb]
  \centering
  \begin{adjustbox}{max width=\linewidth}
    \includegraphics[height=0.5\textheight,keepaspectratio]{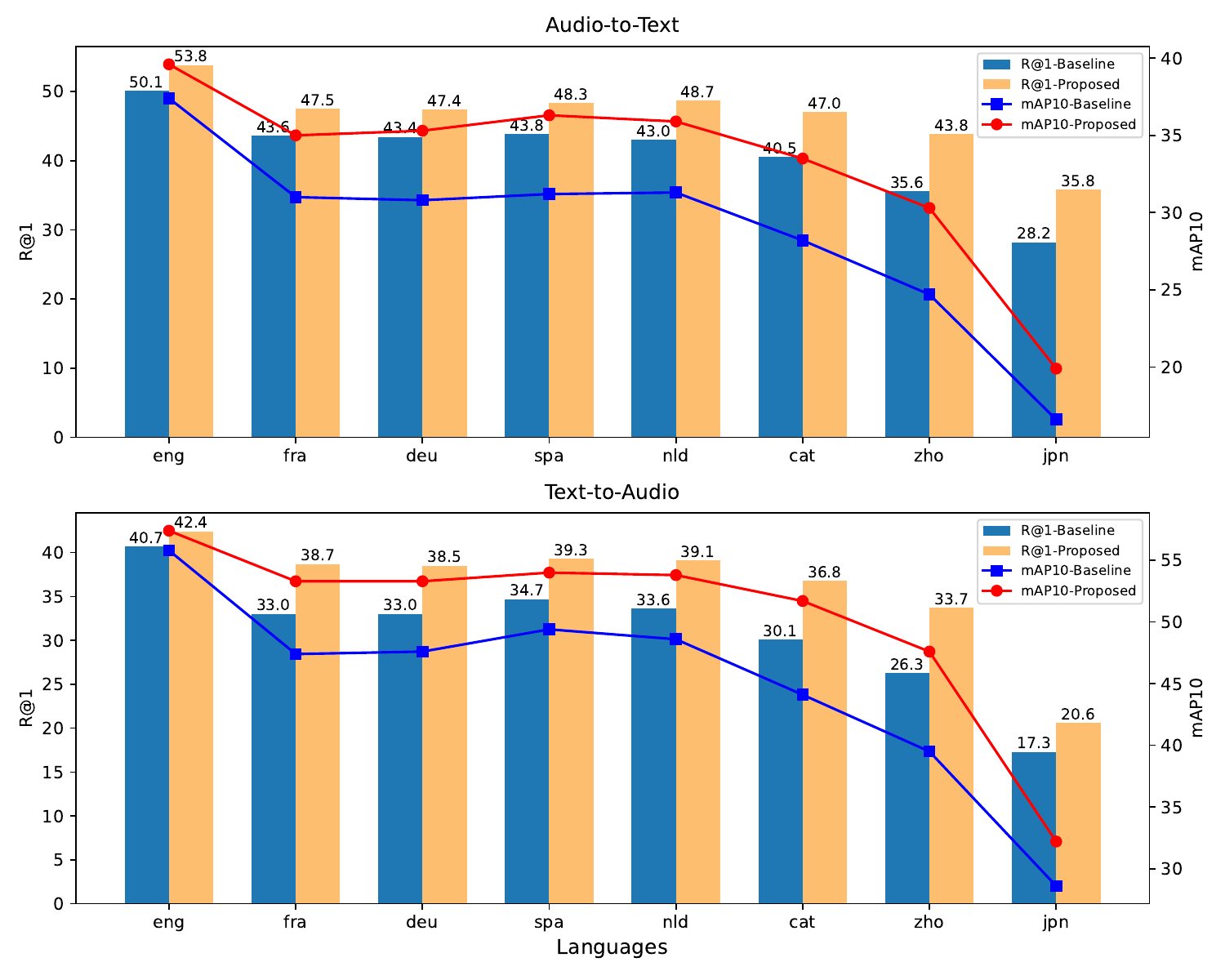}
  \end{adjustbox}
  \caption{ 
  Multilingual evaluation results on AudioCaps, where the x-axis represents the tested target language, with translations obtained by ChatGPT.
  The baseline model represents training on the original, English captions, whereas ``proposed'' represents using mixture LE.
  These observations are consistent with Clotho.
  }
  \label{fig:multilingual-test}
\end{figure}

\begin{table*}[bt]
\centering
\caption{A comparison between our proposed method against previous approaches on English test sets of AudioCaps and Clotho.
Results in {\color{xiaomi_gray} gray} represent the multimodal model.
For all results, higher is better and best results are highlighted in bold.}
\resizebox{\textwidth}{!}{
\begin{tabular}{lcccc|ccc|ccc|ccc}
\toprule
\multirow{4}{*}{Model} & \multirow{4}{*}{Training Type} & \multicolumn{6}{c}{AudioCaps} & \multicolumn{6}{c}{Clotho} \\
\cmidrule(lr){3-14}
& & \multicolumn{3}{c}{Audio-to-Text} & \multicolumn{3}{c}{Text-to-Audio} & \multicolumn{3}{c}{Audio-to-Text} & \multicolumn{3}{c}{Text-to-Audio} \\
\cmidrule(lr){3-14}
& & R@1 & R@5 & R@10 & R@1 & R@5 & R@10 & R@1 & R@5 & R@10 & R@1 & R@5 & R@10 \\
\midrule
        CLAP-HTSAT~\cite{deshmukh2022audio} & \multirow{13}{*}{Pretraining} & 41.9 & 73.1 & 84.6 & 34.6 & 70.2 & 82.0 & 20.0 & 44.9 & 58.7 & 16.7 & 41.1 & 54.1 \\
        LAION~\cite{wu2023largescale} & & 45.8 & 80.9 & 91.6 & 36.1 & 71.8 & 83.9 & 25.7 & 51.5 & 63.4 & 18.2 & 42.5 & 54.4 \\
        LAION (fusion)~\cite{wu2023largescale} & & 45.8 & 80.9 & 91.6 & 35.1 & 71.5 & 83.6 & 25.7 & 51.5 & 63.4 & 18.2 & 42.5 & 54.4 \\
        CNN14-BERT~\cite{mei2023wavcaps} & & 44.6 & 76.3 & 86.2 & 34.7 & 69.1 & 82.5 & 25.9 & 52.6 & 65.8 & 21.2 & 46.4 & 59.4 \\
        HTSAT-BERT~\cite{mei2023wavcaps} & & 51.7 & 82.3 & 90.6 & 39.7 & 74.5 & 86.1 & 23.4 & 50.9 & 63.4 & 19.5 & 45.2 & 58.2 \\
        HTSAT-22+GPT2~\cite{elizalde2024natural} & & 42.5 & - & - & 35.6 & - & - & 22.9& - & - & 15.7 & - & - \\
        FLAP~\cite{yeh2023flap} & & 51.5  &  82.5   & 92.5 & 40.4 &  74.7 &  85.0 & 21.6  &  51.2  & 63.1 & 17.4 &  41.3&  53.7 \\
        FLAP (fusion)~\cite{yeh2023flap} & & 53.0  &  84.1   & 92.6 & 41.5 &  75.5  & 86.0 & 25.5  &  53.4  & 67.9 & 20.3 &  46.5  & 58.8 \\
        BLAT~\cite{xu2024blat} & & 40.4 &  -  & 85.7 & 33.3 & - & 82.4 & 13.9 & - & 48.2 & 12.3 & - & 46.1 \\
        OnePeace~\cite{wang2023onepeace} &  & {\color{xiaomi_gray}51.0} & {\color{xiaomi_gray}81.9} & {\color{xiaomi_gray}92.0} & {\color{xiaomi_gray}42.5} & {\color{xiaomi_gray}77.5} & {\color{xiaomi_gray}88.4} & {\color{xiaomi_gray}27.1} & {\color{xiaomi_gray}52.3} & {\color{xiaomi_gray}65.4} & {\color{xiaomi_gray}22.4} & {\color{xiaomi_gray}49.0} & {\color{xiaomi_gray}62.7} \\
        Cacophony~\cite{zhu2024cacophony} & & 55.3 &  83.6  & 92.4 & 41.0 & 75.3 & 86.4 & 26.5 & 54.1 & 67.3 & 20.2 & 45.9 & 58.8 \\
        CED+BERT & & 52.0 & 84.0 & 91.3 & 39.0 & 75.3 & 87.3 & 28.0 & 55.8 & 70.4 & 23.1 & 50.0 & 64.3 \\
        {Proposed} & & 55.7 & 81.9 & 90.8 & 40.4 & 75.4 & 87.1 & 29.3 & 53.6 & 68.0 & 23.6 & 50.9 & 64.9 \\
        \midrule
        CNN14-NetRVLAD~\cite{lou2022audiotext} & \multirow{6}{*}{Fine-tuning} & 33.3 & 67.6 & 80.6 & 29.3 & 65.2 & 79.3 & 13.0 & 32.9 & 45.4 & 13.1 & 33.1 & 45.1 \\
        BLAT~\cite{xu2024blat} & & 47.5 &  -  & 87.6 & 38.2 & - & 85.1 & 17.9 & - & 50.9 & 13.7 & - & 48.9 \\
        CNN14-BERT~\cite{mei2023wavcaps} & & 45.7 & 76.1 & 87.7 & 35.1 & 70.0 & 82.1 & 27.1 & 52.7 & 66.3 & 21.5 & 47.9 & 61.9 \\
        HTSAT-BERT~\cite{mei2023wavcaps} & & 54.6 & 85.2 & 92.4 & 42.2 & 76.5 & 87.1 & 26.9 & 52.6 & 64.9 & 19.7 & 45.7 & 59.4 \\
        {Proposed} & & 59.3 & 86.3 & 94.0 & 45.6 & 81.0 & \textbf{90.5} & 30.5 & \textbf{58.4} & \textbf{70.7} & 24.7 & 53.6 & \textbf{67.0} \\
        \quad + mixture LE & & \textbf{60.7} & \textbf{86.9} & \textbf{94.8} & \textbf{45.9} & \textbf{81.3} & 90.2 & \textbf{30.9} & 57.5 & 70.2 & \textbf{25.0} & \textbf{53.7} & 66.6 \\
\bottomrule
\end{tabular}
}
\label{tab:sota_results}
\end{table*}


\subsection{Multilingual Capabilities}
\label{ssec:multilingual}
In \Cref{ssec:evaluate_LE}, training with the mixture LE approach equips the model with multilingual audio-text retrieval capabilities. 
It yields improved performance on multilingual test sets compared to the base model 
trained solely on English captions, as depicted in \Cref{fig:multilingual-test}.
Performance for most languages noticeably improves across all tested languages.
However, the retrieval performance for Japanese is suboptimal, primarily due to the complexity of the Japanese text encoder's tokenizer. 
Future work may involve adjusting the proportion of Japanese data in the training set to enhance the existing language ratios.

\subsection{Comparison against previous works}

In \Cref{tab:sota_results}, we compare our proposed approach against previous methods for audio-text retrieval on English test sets. 

During the pretraining with the large dataset, the CED model shows a significant improvement in modeling variable-length audio (Clotho test set) compared to HTSAT, utilizing the same BERT-based text encoder. 
Additionally, the utilization of the SONAR text encoder further enhances the audio-text retrieval performance, demonstrating superior overall average performance compared to the current SOTA models.
Notably, our work outperforms previous approaches that utilized additional training data on Clotho, by a significant margin. 


With only pretraining, our results on Text-to-Audio slightly underperform in terms of R@1 against previous approaches.
However, in terms of Audio-to-Text performance, our approach largely outperforms previous attempts. 
Upon fine-tuning, substantial performance gains are observed, particularly notable in the AudioCaps test set. 
The mixture LE also contributes to enhanced performance during the fine-tuning phase, with most metrics on both the AudioCaps and Clotho test sets reaching the SOTA.
This comparative analysis against SOTA models highlights the efficacy of the proposed approach in modeling audio and text relationships across languages. 
The findings underscore the potential benefits of leveraging multilingual data and advanced text encoders for developing robust audio-text retrieval systems.

\subsection{Ablation studies}
\label{ssec:ablation}


\begin{table}[tb]
\centering
\caption{The performance of mixture LE on the English test set under different data mixing ratios.}
\resizebox{\columnwidth}{!}{
\begin{tabular}{p{0.5cm}cc|cc|cc|cc}
\toprule
\multirow{4}{*}{mix} & \multicolumn{4}{c}{AudioCaps} & \multicolumn{4}{c}{Clotho} \\
\cmidrule(lr){2-9}
& \multicolumn{2}{c}{Audio-to-Text} & \multicolumn{2}{c}{Text-to-Audio} & \multicolumn{2}{c}{Audio-to-Text} & \multicolumn{2}{c}{Text-to-Audio} \\
\cmidrule(lr){2-9}
& R@1 & mAP10 & R@1 & mAP10 & R@1 & mAP10 & R@1 & mAP10 \\
\midrule
10\% & 53.8 & 39.6 & 42.4 & 57.4 & 24.6 & 16.4 & 18.9 & 29.9 \\
20\% & 53.3 & 40.3 & 41.8 & 57.4 & 23.2 & 16.1 & 19.8 & 31.1 \\
30\% & 52.3 & 39.6 & 42.8 & 57.8 & 23.4 & 16.0 & 19.0 & 30.0 \\
40\% & 51.5 & 39.1 & 41.4 & 56.7 & 22.7 & 15.3 & 18.2 & 29.3 \\
50\% & 51.1 & 38.5 & 41.5 & 56.5 & 22.7 & 15.3 & 18.4 & 29.7 \\
\bottomrule
\end{tabular}
}
\label{tab:multilingual_percent_results}
\end{table}

We explore the impact of varying LE mixing ratios on the performance of the English test set during training, as shown in \Cref{tab:multilingual_percent_results}. Our experimental findings suggest using mixing ratios between 10\% and 30\%, with 10\% adopted in our experiments. Beyond 30\%, it adversely affects the model's mAP10 performance. This is primarily due to the utilization of the same audio for multilingual audio-text pairs but with different text captions. 
A higher mixing ratio results in a greater number of text captions per audio. 
For instance, at a 40\% mixing ratio, one English caption plus seven additional languages equates to an average of 3.8 text captions per audio. 
This increased complexity poses challenges for contrastive learning. 
When incorporating more language categories, we recommend reducing the mixing ratio to mitigate its impact on the multilingual audio-text retrieval model.

\section{Conclusion}

In this work, we introduce LE, a simple text augmentation approach for audio-text retrieval, aiming to enable multilingual audio-text retrival. 
We showcase the effectiveness of employing both single-language and mixed-language enhancement for this task.
The results on the English caption test set demonstrate significant improvements, laying a strong foundation for multilingual audio-text retrieval. 
Our exploration across various languages yields promising outcomes, with the incorporation of the mixture LE achieving SOTA results. 
This model also exhibits robust multilingual retrieval capabilities, enhancing its utility for real-world applications.



\bibliographystyle{IEEEtran}
\bibliography{mybib}

\end{document}